\documentclass{article}
\usepackage{graphicx}
\usepackage{amsmath}
\usepackage{color}

\begin{document}

\title{A possible experimental check of the uncertainty relations by means of
homodyne measuring photon quadrature}
\author{V.I. Man'ko$^a$, G. Marmo$^b$, A. Simoni$^b$, F. Ventriglia$^b$ \\
{\footnotesize \textit{$^a$P.N.Lebedev Physical Institute, Leninskii
Prospect 53, Moscow 119991, Russia }}\\
{\footnotesize {(e-mail: \texttt{manko@na.infn.it})}}\\
\textsl{{\footnotesize {$^b$Dipartimento di Scienze Fisiche dell'
Universit\`{a} ``Federico II" e Sezione INFN di Napoli,}}}\\
\textsl{{\footnotesize {Complesso Universitario di Monte S. Angelo, via
Cintia, 80126 Naples, Italy}}}\\
{\footnotesize {(e-mail: \texttt{marmo@na.infn.it, simoni@na.infn.it,
ventriglia@na.infn.it})}}}
\maketitle

\begin{abstract}
We suggest to use the photon homodyne detection experimental data for
checking the Heisenberg and Schr\"{o}dinger-Robertson uncertainty relations,
by means of measuring optical tomograms of the photon quantum states.
\end{abstract}

\section{ Uncertainty relations}

In quantum mechanics and quantum optics a key role in distinguishing
classical and quantum domains is played by the uncertainty relations of
Heisenberg \cite{Heis27} and Schr\"{o}dinger-Robertson \cite{Shr30,Rob29}.
The aim of this work is to suggest a direct experimental check of the
uncertainty relations and control accuracy of the experiments of photon
homodyne detection \cite{Yuen70}. The photon homodyne quadrature was
measured in \cite{Raymer} and in a series of later experiments, see, e.g.,
\cite{Mlynek,Lvovski,Solimeno}. These experiments had the goal to measure
the photon quantum state described by a Wigner function \cite{Wig32}.

As it was shown in \cite{BerBer,VogelRis}, the Wigner function can be
reconstructed if one knows the optical tomogram of the quantum state. In the
experiments \cite{Raymer,Mlynek,Lvovski,Solimeno} with the measuring of the
photon homodyne quadrature by a homodyne detector, the output of the
experimental result was the photon state optical tomogram $\mathcal{W}%
(X,\theta )$. This is the non negative normalized distribution function of
the homodyne quadrature $X$ and of the local oscillator phase $\theta $. By
taking the Radon transform \cite{Radon1917} of $\mathcal{W}(X,\theta )$ one
gets the Wigner function of the photon quantum state.

Our aim is to show that the same experimental setup can be used to check the
uncertainty relations using the measured optical tomograms of photon quantum
states. Since quantum uncertainty relations play a key role in the
foundation of quantum mechanics, it seems reasonable to have a direct method
for an experimental check of both the uncertainty relations and the accuracy
degree of the measurements. These can be controlled and, in principle
improved, in experiments with homodyne detection.

The experimental fulfilling of the uncertainty relations of Heisenberg,
\textit{per se}, dos not characterize a genuine quantum mechanical behaviour
\cite{Beppe george olga PL}, see also \cite{deGosson}. In fact, there are
examples of Hermitian trace class operators which fulfill Heisenberg
uncertainty relations when used as quantum density states \cite{Beppe george
olga PL,O'Connell}. Nevertheless these operators are non-positive, so they
cannot represent any quantum state.

In view of this remark, to have direct independent experimental confirmation
of quantum mechanics foundations, one should check experimentally not only
the quantum uncertainty relations for the second moments of position and
momentum (which are either the Heisenberg or the Schr\"{o}dinger-Robertson
uncertainty relations), but also the quantum inequalities, available in
standard quantum mechanics (see, e.g., the review \cite{Dodonov183}).

We suggest in this paper a way to make such confirmation by using homodyne
detection of photon quantum states. In the next section 2, we review the
tomographic probability formulation of quantum mechanics \cite
{Mancini,manman,CSTom,Ventriglia1,PLASemient,Ventriglia3}. In section 3, we
present the Heisenberg and Schr\"{o}dinger-Robertson uncertainty relations
in the tomographic probability representation of quantum states. In section
4, a suggestion of an experimental check of these uncertainty relations is
discussed. Conclusions and perspective are given in section 5.

\section{Tomograms of quantum states}

According to \cite{Mancini} the quantum state of a photon with Wigner
function $W(q,p)$ is described by the Radon transform of the function
\begin{equation}
\mathcal{W}(X,\mu ,\nu )=\int W(p,q)\delta (X-\mu q-\nu p)\frac{dpdq}{2\pi }%
,(\hbar =1)
\end{equation}
\ which is called symplectic tomogram of the state. The tomogram is
nonnegative and satisfies the normalization condition
\begin{equation}
\int \mathcal{W}(X,\mu ,\nu )dX=1.
\end{equation}
The real parameters $\mu $ and $\nu ,$ in the case when $\mu =\cos \theta $
and $\nu =\sin \theta ,$ provide the phase $\theta $ of a local oscillator
in the experiments with homodyne detecting photon states and $\mathcal{W}%
(X,\mu ,\nu )$ becomes the optical tomogram $\mathcal{W}(X,\theta )$ of \cite
{BerBer,VogelRis} which is directly measured in this experiments. For $\mu
=1 $ and $\nu =0$ the symplectic tomogram yields the position probability
distribution in the quantum state (first quadrature probability distribution
in the photon state). For $\mu =0$ and $\nu =1$ one has the momentum (second
photon quadrature) probability distribution. The symplectic and optical
tomograms are connected by an invertible relation
\begin{eqnarray}
\mathcal{W}(X,\theta ) &=&\mathcal{W}(X,\cos \theta ,\sin \theta ) \\
\mathcal{W}(X,\mu ,\nu ) &=&\frac{1}{\sqrt{\mu ^{2}+\nu ^{2}}}\mathcal{W}%
\left( \frac{X}{\sqrt{\mu ^{2}+\nu ^{2}}},\arctan \frac{\nu }{\mu }\right) .
\label{arctang}
\end{eqnarray}
The Wigner function of the photon quantum state is determined by the
symplectic tomogram \cite{Mancini}:
\begin{equation}
W(p,q)=\int \frac{1}{2\pi }\mathcal{W}(X,\mu ,\nu )\exp \left[ i(X-\mu q-\nu
p)\right] dXd\mu d\nu .  \label{Wigner}
\end{equation}
Introducing polar coordinates $\mu =\sqrt{\mu ^{2}+\nu ^{2}}\cos \theta ,\nu
=\sqrt{\mu ^{2}+\nu ^{2}}\sin \theta $ one can reduce Eq. (\ref{Wigner}) to
a standard Radon integral \cite{Radon1917} used for reconstructing the
Wigner function from the experimentally found optical tomogram $\mathcal{W}%
(X,\theta )$ in the aforementioned experiments. In view of the physical
meaning of the optical tomogram one can calculate higher moments of the
probability distribution
\begin{equation}
\left\langle X^{n}\right\rangle (\mu ,\nu )=\int X^{n}\mathcal{W}(X,\mu ,\nu
)dX,n=1,2,...  \label{moments}
\end{equation}
for any value of the parameters $\mu $ and $\nu ,$ in particular for any
given phase of the local oscillator $\theta .$ This provides the possibility
to check the inequalities for the quantum uncertainty relations.

\section{Schr\"{o}dinger-Robertson uncertainty relations}

The Heisenberg uncertainty relation connects position and momentum variances
$\sigma _{QQ}$ and $\sigma _{PP}$ by means of an inequality. In the
tomographic probability representation the Heisenberg relation reads (see,
e. g., \cite{olgaJRLR}):
\begin{eqnarray}
\sigma _{PP}\sigma _{QQ} &=&\left( \int X^{2}\mathcal{W}(X,0,1)dX-\left[
\int X\mathcal{W}(X,0,1)dX\right] ^{2}\right) \times  \label{uncertainty} \\
&&\left( \int X^{2}\mathcal{W}(X,1,0)dX-\left[ \int X\mathcal{W}(X,1,0)dX%
\right] ^{2}\right) \geq \frac{1}{4}.  \notag
\end{eqnarray}
The Schr\"{o}dinger-Robertson uncertainty relation contains the contribution
of the position-momentum covariance $\sigma _{QP}$ and reads
\begin{equation}
\sigma _{QQ}\sigma _{PP}-\sigma _{QP}^{2}\geq \frac{1}{4}.
\label{pos-mom covar}
\end{equation}
In view of Eq. (\ref{moments}), the variance $\sigma _{XX}$ of the homodyne
quadrature $X,$ in terms of the parameters $\mu ,\nu $ and the quadratures
variances and covariance, is
\begin{equation}
\sigma _{XX}(\mu ,\nu )=\mu ^{2}\sigma _{QQ}+\nu ^{2}\sigma _{PP}+2\mu \nu
\sigma _{QP}.
\end{equation}
Then one can get the expression of the covariance in terms of the
tomographic characteristics of the state. Taking $\mu =\nu =\sqrt{2}/2$
corresponding to the local oscillator phase $\theta =\pi /4$ one has
\begin{equation}
\sigma _{QP}=\sigma _{XX}\left( \theta =\frac{\pi }{4}\right) -\frac{1}{2}%
(\sigma _{ QQ}+\sigma _{PP})
\end{equation}
where $\sigma _{PP}$ and $\sigma _{QQ}$ are the factors appearing in the
left hand side of Eq. (\ref{uncertainty}) respectively. The term $\sigma
_{XX}(\theta =\pi /4)$ is given by Eq. (\ref{moments}) as
\begin{equation}
\sigma _{XX}\left( \theta =\frac{\pi }{4}\right) =\left\langle
X^{2}\right\rangle \left( \frac{\sqrt{2}}{2},\frac{\sqrt{2}}{2}\right) -%
\left[ \left\langle X\right\rangle \left( \frac{\sqrt{2}}{2},\frac{\sqrt{2}}{%
2}\right) \right] ^{2}.
\end{equation}

\section{Checking uncertainty relations}

On the base of the obtained formulae we suggest the following procedure to
check the Heisenberg and Schr\"{o}dinger-Robertson uncertainty relations.
First one obtains the function $\mathcal{W}(X,\theta ),$ which is the
optical tomogram, from the standard homodyne detection of a photon state. It
means that one has also the symplectic tomogram $\mathcal{W}(X,\mu ,\nu )$
according to Eq. (\ref{arctang}). Formula (\ref{uncertainty}) can then be
directly checked if one obtains from the experimental data the integrals in
the left-hand side for $\mathcal{W}(X,\theta =0)$ and $\mathcal{W}(X,\theta
=\pi /2)$ and compares the product $\sigma _{XX}(\theta =0)\sigma
_{XX}(\theta =\pi /2)$ with $1/4$. The check of Schr\"{o}dinger-Robertson
uncertainty relations requires extra elaboration of the available
experimentally obtained optical tomogram of photon quantum state. We express
this procedure as the following inequality for optical tomogram. Let us
calculate the function $F(\theta )$ which we call ``tomographic uncertainty
function'':
\begin{eqnarray}
F(\theta ) &=&\left( \int X^{2}\mathcal{W}(X,\theta )dX-\left[ \int X%
\mathcal{W}(X,\theta )dX\right] ^{2}\right) \times  \label{F(Theta)} \\
&&\left( \int X^{2}\mathcal{W}(X,\theta +\frac{\pi }{2})dX-\left[ \int X%
\mathcal{W}(X,\theta +\frac{\pi }{2})dX\right] ^{2}\right)  \notag \\
&&-\left\{ \int X^{2}\mathcal{W}(X,\theta +\frac{\pi }{4})dX-\left[ \int X%
\mathcal{W}(X,\theta +\frac{\pi }{4})dX\right] ^{2}\right.  \notag \\
&&-\frac{1}{2}\left[ \int X^{2}\mathcal{W}(X,\theta )dX-\left[ \int X%
\mathcal{W}(X,\theta )dX\right] ^{2}\right.  \notag \\
&&+\left. \left. \int X^{2}\mathcal{W}(X,\theta +\frac{\pi }{2})dX-\left[
\int X\mathcal{W}(X,\theta +\frac{\pi }{2})dX\right] ^{2}\right] \right\} -%
\frac{1}{4}.  \notag
\end{eqnarray}
The tomographic uncertainty function must be non-negative
\begin{equation}
F(\theta )\geq 0  \label{F ineq}
\end{equation}
for all the values of the local oscillator phase angle $0\leq \theta \leq
2\pi .$ The previous Eq. (\ref{F(Theta)}) for $\theta =0$ yields Eq. (\ref
{pos-mom covar}). Thus, choosing the values $\theta =0,\pi /4,\pi /2$ out
from experimental optical tomogram $\mathcal{W}(X,\theta )$ data, one can
check both the Heisenberg uncertainty relation (Eq.\ref{uncertainty}) and
the inequality (\ref{pos-mom covar}). Moreover one can check also the above
inequality (\ref{F ineq}) by using tomographic experimental data
corresponding to all values of angles $\theta ,\theta +\pi/4,\theta +\pi /2.$

\section{Conclusion}

We point out the main results of this work. We suggest to use the known
experimental data obtained by measuring quantum states by means of optical
tomographic method, which in all the available experiments were used to find
the Wigner function, as a tool to check the quantum uncertainty relations.
Our suggestion consists of elaborating the experimental optical tomogram
data for computing the tomographic uncertainty function $F(\theta )$ defined
in Eq. (\ref{F(Theta)}), instead of using the data, as usual, in a Radon
integral transform leading to the Wigner function. The function of the local
oscillator phase $F(\theta )$ contains integrations for different fixed
values of $\theta .$ In the case of the computation of the Wigner function
the integration over the local oscillator phase is performed. However, even
though the integrals to evaluate differ from the integrals in the Radon
transform, they do not contain extra mathematical complications.

The suggested experimental checking of the quantum uncertainty relations can
be used not only to test the degree of experimental accuracy with which the
uncertainty relations are known today, but also to control the correctness
of the experimental tools used in homodyne detection of photon states.

There exist inequalities in which the higher momenta of quadrature
components are involved (see, e.g., the review \cite{Dodonov183}). One can
reformulate these higher order inequalities in terms of the tomographic
quadrature momenta given in Eq. (\ref{moments}) in order to obtain extra
inequalities, again expressed in terms of the experimental values of the
optical tomogram.

The tomographic probability approach can be applied also for two-mode and
multi-mode photon states, in particular for Gaussian states, whose
properties, like photon statistics, are sufficiently known (see, e.g., \cite
{OlgaPhysRev,MatteoParisBook}).

The tomographic entropic uncertainty relations which are associated with
position and momentum probability distributions were discussed in \cite
{margarita}.

\noindent\textbf{Acknowledgements} V.I.M. thanks I.N.F.N. and the University
``Federico II'' of Naples for their hospitality, the Russian Foundation for
Basic Research under Projects Nos.~07-02-00598 and 08-02-90300, and the
Organizers of the IV workshop \textit{in memoriam} of Carlo Novero
``Advances in Foundations of Quantum Mechanics and Quantum Information with
atoms and photons'' for invitation and kind hospitality.

\end{document}